\documentclass{eptcs}
\usepackage{breakurl}             

\title{On Varieties of Automata Enriched\\ with an Algebraic Structure\\
(Extended Abstract)}
\author{Ond\v rej  Kl\'{i}ma
\institute{Department of Mathematics and Statistics,\\ Masaryk University,\\ 
Brno, Czech Republic 
\thanks{Supported by the project
``Algebraic Methods in Quantum Logic''
by ESF, No. CZ.1.07/2.3.00/20.0051.} }
\email{klima@math.muni.cz} 
}
\def\titlerunning{On Varieties of Automata Enriched with an Algebraic Structure}
\def\authorrunning{Ond\v rej  Kl\'{i}ma}
\begin{document}
\maketitle

\begin{abstract}
Eilenberg correspondence, based on the concept of syntactic monoids, 
relates varieties of regular languages with pseudovarieties of finite monoids. 
Various modifications of this correspondence related more general classes 
of regular languages with classes of more complex algebraic objects.
Such generalized varieties also have natural counterparts formed by
classes of finite automata equipped with a certain additional algebraic structure.  
In this survey, we overview several variants of such varieties of enriched automata. 
\end{abstract}


Algebraic theory of regular languages is a well established field in the theory of formal languages.
A basic ambition of this theory is to obtain 
effective characterizations of various natural classes of regular languages.
The fundamental concept is the notion of {\em syntactic monoid} of a given regular language $L$,
which is the smallest possible monoid recognizing the language $L$, and which is isomorphic
to the transition monoid of the minimal automaton of $L$.
First examples of natural classes of languages, which were effectively characterized 
by properties of syntactic monoids, were the star-free languages~\cite{schutz} having
aperiodic syntactic monoids and the piecewise testable languages~\cite{simon-pw} having
$\mathcal J$-trivial syntactic monoids. 
A~general framework for discovering relationships
between properties of regular languages and properties of monoids
was provided by Eilenberg~\cite{eilenberg}, who established a one-to-one
correspondence between so-called {\em varieties} of regular languages
and \emph{pseudovarieties} of finite monoids.
Here varieties of languages are classes closed under taking quotients, preimages under morphisms and Boolean operations.
On the other hand pseudovarieties of finite monoids are classes closed under taking finite direct products, 
submonoids and morphic images.
Thus a membership problem for a given variety of regular languages can be translated to a membership problem
for the corresponding pseudovariety of finite monoids.
An advantage of this translation is that pseudovarieties of monoids are exactly classes
of finite monoids which have equational description by pseudoidentities~\cite{reiterman}.

The goal of this contribution is not to overview all notions and applications of the 
algebraic theory of regular languages. For thorough introduction
to that theory we refer to~\cite{pin-handbook}. Other overviews are for example
\cite{pin-survey} and \cite{weil-survey}.
A more detailed information concerning  the theory of pseudovarieties of finite monoids can be 
found in the survey~\cite{almeida-survey} or in the books~\cite{almeida} and~\cite{steinberg}.

We should mention that many interesting classes of regular languages, 
which are studied by the algebraic methods, come from logic. 
It is well known that regular languages which are definable in
the first order logic of finite linear orderings are exactly star-free languages~\cite{McNaughton}.
Within the class of star-free languages, there were defined the so-called
\emph{dot-depth hierarchy}~\cite{dot-hierarchy}  and closely related 
\emph{Straubing--Th\' erien hierarchy}~\cite{str-hierarchy,therien-hierarchy}.
In~\cite{thomas} it was shown that a language belongs to the $n$th level
of the latter hierarchy if and only if it is definable by a formula
with $n$~alternations of quantifiers.
Moreover, the class of star-free languages is exactly  the class of all languages definable by 
linear temporal logic~\cite{kamp}. For a recent survey on the classes of languages
given  by fragments of first-order logic we refer to~\cite{diekert-survey} and~\cite{str-weil}.
Some recent results can be found for example in~\cite{weil-logic} and~\cite{zeitoun-separ}.

Since not every natural class of languages is closed under all mentioned operations,
various generalizations of the notion of varieties of languages were studied. 
One possible generalization is the notion of {\em positive varieties} of languages introduced 
in~\cite{pin-positive} for which an equational characterization was given in~\cite{pin-weil};
the positive varieties need not be closed under complementation.
In the same direction one can consider varieties which need not be closed under taking unions
(see~\cite{polak-brno}).
We shall return to  these concepts later. 
Another possibility is to weaken the closure property for preimages under morphisms.
In this way one can consider $\mathcal C$-varieties of regular languages
which were introduced in~\cite{straubing} and whose equational description was given in~\cite{michal}.
Here we require the presence of preimages under morphisms only for morphisms from a certain special class $\mathcal C$.
An important example is the class formed by morphisms which map letters to letters; 
such varieties of languages
(so-called {\em literal varieties)} and the corresponding pseudovarieties of monoids with marked generators
(so-called {\em monoid-generator pairs}) were studied in~\cite{esik-larsen}.
Classes of languages with a complete absence of the preimages requirement were studied   
in~\cite{pin-lattices}.

In our contribution we would like to consider varieties of automata as another
natural counterpart to varieties of regular languages. 
We should emphasize that the considered automata are deterministic finite automata.
Characterizing of varieties of languages by properties of minimal automata
is quite natural, since usually we assume that an input of a membership problem
for a fixed variety of languages is given exactly by a minimal deterministic automaton.
For example, if we want to effectively test whether an input language is piecewise testable,
we do not want to compute its syntactic monoid which could be quite large\footnote{More than 
$(n-1)!$ where $n$ is the number of states
of the minimal automaton, see~\cite{broz-j-triv} for precise bounds.}.
Instead of that we consider a condition which must be satisfied by its minimal automaton
and which was given in the original Simon's paper~\cite{simon-pw}.
This characterization was used in~\cite{stern} and~\cite{trahtman}
to obtain a polynomial and quadratic algorithm, respectively, for testing piecewise testability.
In~\cite{dlt13-klima} Simon's condition was reformulated and 
the so-called {\em locally confluent acyclic automata}\footnote{These automata recognize exactly
piecewise testable languages and paper~\cite{dlt13-klima} contains 
a new (purely automata based) proof which does not use Simon's original result.} 
were defined.
Therefore we are looking for a general definition of a term {\em variety of automata},
to obtain a setting in which we could talk, for example, about the variety of locally confluent acyclic automata. 

Let us consider a minimal automaton $\mathcal A_L$ of a regular language $L$.
A first easy observation is the following: if we change the initial state in $\mathcal A_L$ then
the resulting automaton recognizes a left quotient of the original language $L$. 
Similarly but not trivially,
if we change the final states, the resulting automaton recognizes a Boolean combination of right quotients 
of the original language $L$. Since we are interested in characterizations of varieties of languages, 
the choice of an initial state and final states can be left free and 
we can consider only underlying labeled graphs\footnote{Such automata without initial and final states 
are sometimes called semiautomata in the literature.} which will form our varieties of automata.
Furthermore, since varieties of languages are closed under taking unions and intersections, we need to include 
direct products 
of automata in our varieties of automata. 
Considering a preimage of a given regular language $L$ under some morphism $f$, one can construct 
an automaton from the minimal automaton $\mathcal A_L$ of $L$, so-called {\em $f$-subautomaton}, where
states form a subset and a new action by each letter $a$ is the same as the 
action by the word $f(a)$ in the original automaton.
Since these constructions generate new automata, namely products of automata
and $f$-subautomata, and since we are mainly interested in minimal automata, we also include into our variety
of automata all morphic images of existing automata.
Finally, from technical reasons we add disjoint unions of automata.
Thus a variety of automata will be a class of automata closed under taking 
products, disjoint unions, morphic images and $f$-subautomata.
And of course, when we are limited to morphisms from a certain class $\mathcal C$,
we can even talk about $\mathcal C$-varieties of automata. 
Then one can prove an Eilenberg type correspondence: varieties of languages correspond to 
varieties of automata. This concept occurred 
in~\cite{esik-ito} in the case of literal morphisms and in~\cite{pin-str} 
under the name varieties of $\mathcal C$-actions. 
In particular, one can consider the variety of all 
counter-free automata~\cite{McNaughton} characterizing star-free languages
or the variety of all locally confluent acyclic automata.

Now we enrich automata by an algebraic structure.
If we start with a deterministic automaton where all states are reachable from the initial one
then we can assign to each state $q$ the set $L_q$ consisting
of all words which are acceptable if the computation starts from this state.
Sometimes $L_q$ is called the {\em future} of the state $q$.
It is known~\cite{broz-minimal} that identifying the states with the same future 
produces a minimal automaton.
Thus a state $q$ in the minimal automaton $\mathcal A_L$ can be identified with its future $L_q$
and therefore it is a subset of $A^*$.  Then such states are ordered by inclusion,
which means that each minimal automaton is implicitly equipped with a partial order.
Moreover, final states\footnote{A state is final if and only if it contains the empty word.} 
form an upward closed subset. This leads to a notion of partially ordered automata where 
actions by letters are isotone mappings and 
languages are recognized by final states which form an upward closed subset.

Furthermore, varieties, or more generally $\mathcal C$-varieties, of 
partially ordered automata\footnote{There exist several papers which use the term ordered 
(deterministic, non-deterministic or two-way) automaton
in a different meaning, e.g. in~\cite{therien-32} it is required that an action by a letter is increasing
but need not be isotone.} can be defined once again as classes which are closed under taking products, disjoint
unions, morphic images and $f$-subautomata. Now one can prove that these varieties of partially ordered automata
correspond to positive varieties of languages. A well known example is 
the level $1/2$ in the Straubing-Th\'erien hierarchy of star-free languages. 
The effective characterization of the level $1/2$ can be found
in~\cite{arfi}. This characterization can be equivalently stated as validity of the identity $1\le x$
in the syntactic ordered monoid of a language~\cite{pin-pulky}.  
Therefore, the corresponding variety of partially ordered automata is formed by automata where
actions are increasing mappings (for a state $q$ and a letter $a$ we have $q\cdot a \ge q$).

Now we return to the representation of the minimal automaton $\mathcal A_L$ of a regular language $L$
where a state $q=L_q$ is a subset of $A^*$ and we consider all possible intersections of states.
Since we have only finitely many states in $\mathcal A_L$,  we 
obtain finitely many intersections. The resulting meet-semilattice $\mathcal S_L$
can be naturally equipped with actions by letters: 
applying a letter $a$ to an 
intersection $\bigcap_{i\in I} q_i$, i.e. a state in 
$\mathcal S_L$, is the intersection of all states $q_i\cdot a$.
If we use as final states those which contain the empty word, then
final states form a principal filter in the semilattice  $\mathcal S_L$.
This idea leads to a notion of a {\em meet automaton} which was introduced in~\cite{meet-automata}.
Here the corresponding varieties of languages are not closed under taking unions, since
in the product of automata the corresponding set of final states is not a principal filter.
Therefore the  corresponding classes of regular languages are 
conjunctive varieties which were defined in~\cite{polak-brno}.
We have already mentioned that the syntactic (ordered) monoid of a language
is isomorphic to the transition (ordered) monoid of the (ordered) minimal automaton of the language.
Analogous statement is valid in the case of meet automata. In particular,
the {\em canonical meet automaton} $\mathcal S_L$ of a language $L$
is a  minimal meet automaton of a given language. Moreover, 
its transition structure is a {\em syntactic semiring} which is a minimal semiring recognizing 
the language and which can be defined analogously to a syntactic monoid (see~\cite{polak-brno}).
In the paper~\cite{meet-automata} there are mentioned some examples of $\mathcal C$-varieties of 
languages which can be characterized via varieties of meet automata.
There is also a close connection between the notion of a canonical meet automaton 
and a notion of a {\em universal automaton} which contains all minimal non-deterministic automata of 
a given regular language
(see~\cite{polak-universal} and~\cite{lombardy}).

One can make one step further. As we add intersections to the representation of minimal automaton,
we can try to add also unions. In other words, we consider the sublattice of the lattice $2^{A^*}$
generated by $\mathcal A_L$.
Since this lattice is distributive, we define an abstract notion of
a {\em distributive lattice automata} ({\em DL-automata}) which are automata enriched 
by a distributive lattice structure, where both operations are compatible with actions by letters. 
Note that this model differs from
lattice automata defined in~\cite{lattice-automata}. We want to define varieties of $DL$-automata
as a natural counterpart of generalized varieties of languages 
which are not required to be closed under taking any of Boolean operations. 
Indeed, such classes naturally occur in the theory of formal languages: 
for example, many classes defined by models of quantum automata are of this kind. 
The goal is a characterization of such classes.
Note that it is also possible to extend this principle, 
consider the Boolean subalgebra of $2^{A^*}$ generated  by $\mathcal A_L$
and define a notion of a {\em  $BA$-automaton}.
Before developing this theory we prefer to clarify all aspects of the theory 
of $DL$-automata, since there are some difficulties.
For example, in the case of meet automata, since actions by letters are morphism 
with respect to the meet operation, 
actions by sets of letters are also morphisms with respect to this operation.
In the case of $DL$-automata, such an extension is not valid.

At the end we could mention that
one can extend the construction in at least two natural directions.
First, the theory of tree languages is a field where many fundamental ideas from the theory of deterministic 
automata were successfully generalized. 
Another recent notion of biautomata (see~\cite{biautomaty} and \cite{ncma13-holzer})
is based on considering both-sided quotients instead of left quotients only.
In both cases one can try to apply the previous constructions
and consider varieties of automata (enriched by an algebraic structure). 
Some papers in this direction already exist~\cite{esik-ivan}.

\subsection*{Acknowledgement}

I would like to express my gratitude to my colleagues Michal Kunc and Libor Pol\'{a}k
for our numerous interesting discussions on the topic.

\nocite{*}
\bibliographystyle{eptcs}
\bibliography{klima}
\end{document}